\documentclass[aps,prl,reprint,superscriptaddress,showpacs,longbibliography,footnoteinbib]{revtex4-2}
\usepackage[latin9]{inputenc}
\usepackage{color}
\usepackage{bm}
\usepackage{amsmath}
\usepackage{amssymb}
\usepackage{graphicx}
\usepackage[unicode=true,
 bookmarks=false,
 breaklinks=false,pdfborder={0 0 1},backref=false,colorlinks=true]
 {hyperref}
\hypersetup{pdftitle={Title},
 plainpages=false,pdfpagelabels,linkcolor=blue,urlcolor=blue,citecolor=blue,pdfdisplaydoctitle=true,pdfduplex=DuplexFlipLongEdge}

\makeatletter
\usepackage{units}\usepackage{wasysym}

\definecolor{orange}{rgb}{0.50, 0.20, 0.0}

\usepackage{bm}
\usepackage{braket}

\makeatother

\begin{document}
\noindent %
\noindent\begin{minipage}[t]{1\columnwidth}%
\global\long\def\ket#1{\left| #1\right\rangle }%
\global\long\def\bra#1{\left\langle #1 \right|}%
\global\long\def\kket#1{\left\Vert #1\right\rangle }%
\global\long\def\bbra#1{\left\langle #1\right\Vert }%
\global\long\def\braket#1#2{\left\langle #1\right. \left| #2 \right\rangle }%
\global\long\def\bbrakket#1#2{\left\langle #1\right. \left\Vert #2\right\rangle }%
\global\long\def\av#1{\left\langle #1 \right\rangle }%
\global\long\def\tr{\text{tr}}%
\global\long\def\Tr{\text{Tr}}%
\global\long\def\pd{\partial}%
\global\long\def\im{\text{Im}}%
\global\long\def\re{\text{Re}}%
\global\long\def\sgn{\text{sgn}}%
\global\long\def\Det{\text{Det}}%
\global\long\def\abs#1{\left|#1\right|}%
\global\long\def\up{\uparrow}%
\global\long\def\down{\downarrow}%
\global\long\def\vc#1{\mathbf{#1}}%
\global\long\def\bs#1{\boldsymbol{#1}}%
\global\long\def\t#1{\text{#1}}%
\end{minipage}
\title{Disorder-induced instability of a Weyl nodal loop semimetal towards
a diffusive topological metal with protected multifractal surface
states}
\author{João S. Silva}
\affiliation{Centro de Física das Universidades do Minho e Porto, LaPMET, Departamento
de Física e Astronomia, Faculdade de Ciências, Universidade do Porto,
4169-007 Porto, Portugal}
\author{Miguel Gonçalves}
\affiliation{CeFEMA, LaPMET, Instituto Superior Técnico, Universidade de Lisboa,
Av. Rovisco Pais, 1049-001 Lisboa, Portugal}
\author{Eduardo V. Castro}
\affiliation{Centro de Física das Universidades do Minho e Porto, LaPMET, Departamento
de Física e Astronomia, Faculdade de Ciências, Universidade do Porto,
4169-007 Porto, Portugal}
\affiliation{Beijing Computational Science Research Center, Beijing 100193, China}
\author{Pedro Ribeiro}
\affiliation{CeFEMA, LaPMET, Instituto Superior Técnico, Universidade de Lisboa,
Av. Rovisco Pais, 1049-001 Lisboa, Portugal}
\affiliation{Beijing Computational Science Research Center, Beijing 100193, China}
\author{Miguel A. N. Araújo}
\affiliation{Beijing Computational Science Research Center, Beijing 100193, China}
\affiliation{CeFEMA, Instituto Superior Técnico, Universidade de Lisboa, Av. Rovisco
Pais, 1049-001 Lisboa, Portugal}
\affiliation{Departamento de Física, Universidade de Évora, P-7000-671, Évora,
Portugal}
\begin{abstract}
Weyl nodal loop semimetals are gapless topological phases that, unlike
their insulator counterparts, may be unstable to small perturbations
that respect their topology-protecting symmetries. Here, we analyze
a clean system perturbed by chiral off-diagonal disorder using numerically
exact methods. We establish that the ballistic semimetallic phase
is unstable towards the formation of an unconventional topological
diffusive metal hosting topological multifractal surface states. Although,
as in the clean case, surface states are exponentially localized along
the direction perpendicular to the nodal loop, disorder induces a
multifractal structure in the remaining directions. Surprisingly,
the number of these states also increases with a small amount of disorder.
Eventually, as disorder is further increased, the number of surface
states starts decreasing.
\end{abstract}
\maketitle
The topological properties of quantum matter are robust to small perturbations,
 such as weak disorder. As long as the clean limit symmetries are
preserved, bulk topological characteristics and associated boundary
states survive until a critical disorder threshold is reached, where
a topological transition takes place.   Particularly interesting examples
of these phases are first \citep{ChiuRMP2016,PhysRevB.106.045417}
and higher order \citep{Li2020,Benalcazar2022} topological insulators,
which, in certain cases, only require disorder to preserve the clean
limit symmetries on average \citep{Song2021},  rendering them particularly
robust and appealing for applications.

Recently, topological semimetals stood out as an important class of
three-dimensional (3D) topological materials not requiring a gap \citep{Armitage2018,annurev-matRes-2019,PhysRevB.96.201401,PhysRevX.8.031076}.
Weyl  nodal loop (WNL) semimetals, in particular, are characterized
by a linearly vanishing density of states (DOS) due to the linear
touch of conduction and valence bands along a loop or line in momentum
space \citep{Yu2017}. Their non-trivial bulk topology gives rise
to characteristic surface states -- the so-called drumhead states
-- which have been detected recently by a variety of methods in different
materials \citep{Bian2016,Belopolski2019,Muechler2019,Sims2019,Hosen2020,Stuart2022}.
This calls for a deeper understanding on the effect of disorder on
the topological phase diagram of these systems.

Interesting consequences of symmetry-breaking disorder where reported
in Refs.~\cite{ShuChen2018,Goncalves2020}. However,  the case of
disorder that preserves the underlying symmetry, such as a WNL semimetal
subjected to chiral symmetric disorder, is particularly intriguing,
and remais open. On one hand, chiral disorder is known to induce an
enhanced DOS at zero energy in chiral symmetric models \citep{Evangelou2003,Garcia-Garcia2006},
which in two-dimensions even becomes logarithmically divergent \citep{GADE1993499},
casting doubts on the stability of the semimetallic phase. On the
other hand, since the symmetry remains unbroken, the robustness of
bulk topological properties and accompanying drumhead states are still
expected, suggesting the stability of the topological semimetal to
weak disorder. In this paper, we solve this apparent contradiction.

\begin{figure}[th]
\begin{centering}
\includegraphics[width=1.\columnwidth]{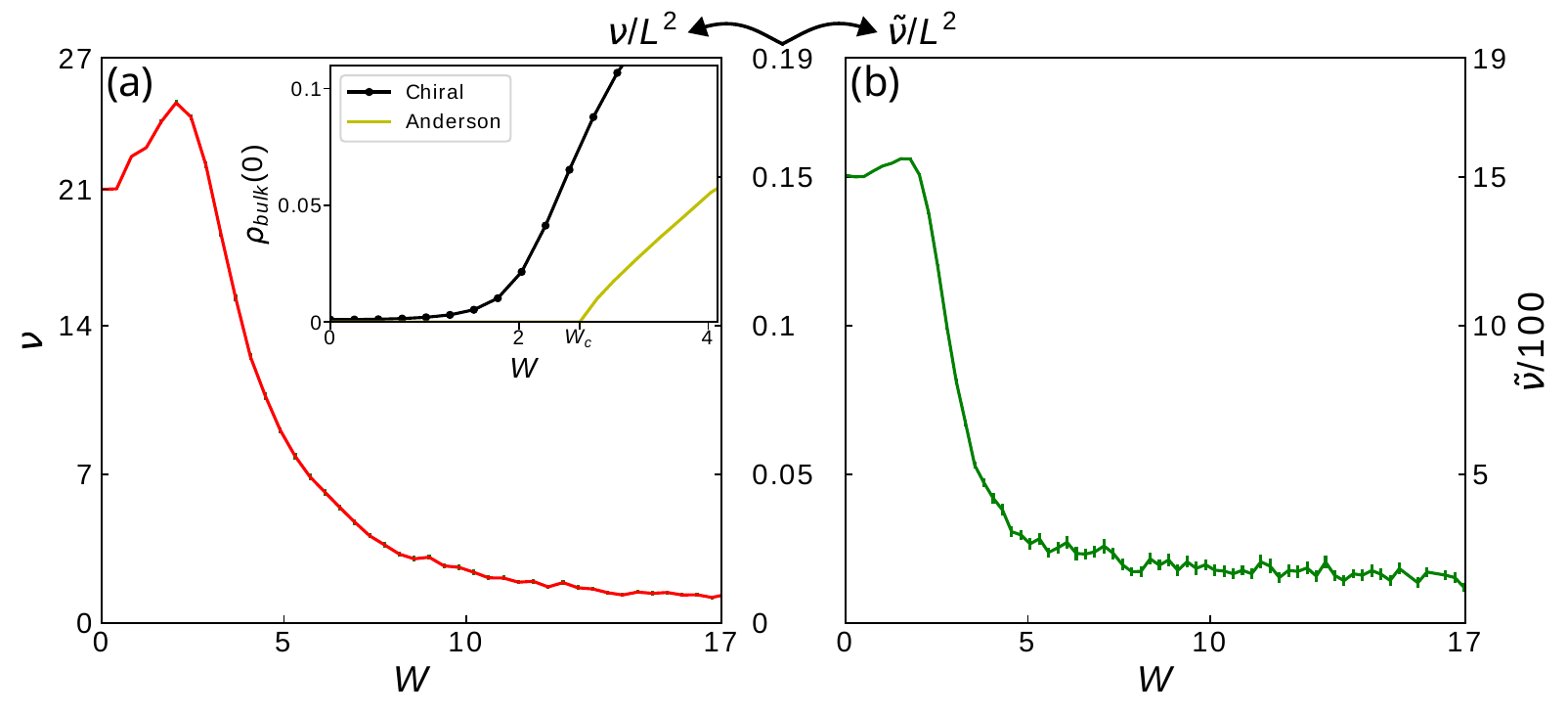} 
\par\end{centering}
\caption{\label{fig:wind-SS-all}(a)~Winding number vs chiral disorder strength
for a system with $L^{3}=12^{3}$ unit cells. The inset shows that
a finite DOS at zero energy appears with increasing disorder, signaling
a metallic phase in contrast with Anderson disorder \citep{Goncalves2020}.
(b)~Estimated number of surface states ($\tilde{\nu}$) versus disorder
obtained with KPM for $L=100$ and $N_{m}=2^{10}$ {[}$L=200$ and
$N_{m}=2^{11}$ for the DOS in the inset of panel~(a){]}.}
\end{figure}

By considering a generic model for a chiral symmetric WNL, we show
that the semimetallic phase is unstable to chiral disorder, and a
finite DOS at zero energy is always present. However, the induced
metallic phase is accompanied by a quantized, nontrivial winding number,
and by zero energy edge states which, for weak disorder, are in higher
number than the clean limit drumhead surface states. The winding number
$\nu$,  which in the clean limit counts the number of $\boldsymbol{k}$
points inside the nodal loop, is shown in Fig.~\ref{fig:wind-SS-all}(a)
versus the disorder strength, $W$. Adding chiral disorder makes this
topological invariant increase, which is accompanied by an increase
of the estimated number of surface states $\tilde{\nu}$, depicted
in Fig.~\ref{fig:wind-SS-all}(b). At the same time, the presence
of a finite zero energy DOS for any finite disorder, as shown in the
inset of Fig.~\ref{fig:wind-SS-all}(a), clearly points to a diffusive
metallic phase \citep{Luo2020}. By further increasing the disorder
strength, both $\nu$ and $\tilde{\nu}$ fade away for general chiral disorder, but a topological transition does not occur.

The presence of non-trivial topology and boundary states together
with a finite bulk DOS at the Fermi energy shows that the WNL is unstable
to a topological metal phase under chiral disorder. However, unlike
previously known topological metals \citep{Ying2018,Bahari2019,Jangjan2021,Xie2021,Cerjan2022,Xie2023},
chiral disorder is an essential ingredient since the system is a semimetal
in the clean limit. Moreover, the surface states of this topological
metal acquire a multifractal structure in real-space, in contrast
to the clean WNL and to models with disorder in one spatial direction \citep{Wang2018}.
This is the first instance of a system hosting topological multifractal
surface states, which realize a new type of bound states in the continuum
\citep{Yang2013,Hsu2016,bic_PhysRevLett.118.166803,bic_PhysRevB.100.075120,Benalcazar2020}.

We consider a two-band model on a cubic lattice with $L^{3}$ cells
and disorder \citep{Goncalves2020,Silva2023}, 
\begin{equation}
\hat{H}=\sum_{\bm{k}}c_{\bm{k}}^{\dagger}H_{\bm{k}}c_{\bm{k}}+\hat{V}\,.\label{eq:total_H}
\end{equation}
The first term describes a clean WNL \citep{Goncalves2020}, with
$\bm{k}$ a 3D Bloch vector, $H_{\bm{k}}=(t_{x}\cos k_{x}+t_{y}\cos k_{y}+\cos k_{z}-m)\tau_{x}+t_{2}\sin k_{z}\tau_{y}$,
with $\tau_{x},\tau_{y}$ Pauli matrices acting on the orbital pseudo-spin
indices $\alpha=1,2$, and $c_{\bm{k}}^{\dagger}=(\begin{array}{cc}
c_{\bm{k},1}^{\dagger} & c_{\bm{k},2}^{\dagger}\end{array})$. In the following, we make the parameter choice $t_{x}=1.1$, $t_{y}=0.9$,
$m=2.12$ and $t_{2}=0.8$. This choice yields a single nodal line,
arising for $k_{z}=0$, with equation $t_{x}\cos k_{x}+t_{y}\cos k_{y}+1-m=0$.
The number of $\bm{k}$ points inside the nodal loop is the number
of drumhead states in a clean WNL \citep{Bian2016,
Yu2015,Kim2015,Weng2015,Li2017,Goncalves2020,Linhu2017,ShuChen2018,Araujobook2021,Silva2023}. The second term is the off-diagonal disorder. In real space, it
reads 
\begin{multline}
\hat{V}=\sum_{\bm{r}}\Bigg[\Bigg(\frac{1}{2}\sum_{\delta=x,y,z}V_{\bm{r}}^{\delta}c_{\bm{r}}^{\dagger}\tau_{x}c_{\bm{r}+\boldsymbol{e}_{\delta}}+{\rm H.c.}\Bigg)\\
+\left.V_{\bm{r}}^{0}c_{\bm{r}}^{\dagger}\tau_{x}c_{\bm{r}}\right],\label{eq:disorder_V}
\end{multline}
where $V_{\bm{r}}^{\delta}=W\omega_{\bm{r}}^{\delta}$, with four
independent random numbers $\omega_{\bm{r}}^{\delta}\in\left[-\frac{1}{2},\frac{1}{2}\right]$,
with $\delta=0,x,y,z$. $W$ measures the disorder strength. The model
in Eq.~(\ref{eq:total_H}) realizes a chiral symmetric disordered
WNL. We later also consider the case of intracell chiral disorder,
where $\omega_{\bm{r}}^{\delta}=0$ for $\delta\neq0$. Unless otherwise
stated, our results are averaged over 100 disorder realizations.

For a disordered chiral system, a winding number, $\nu$, can be calculated
from real space wave functions through the coupling matrix approach
\citep{couplingmatrix,vanderbilt_2018}, by applying twisted boundary
conditions along the $z$ direction. For a given twist angle, $\theta$,
the ground-state many-body wave function for a half filled system,
$\ket{\Phi({\theta})}$, is a Slater determinant of single-particle
states, $|\psi_{j}(\theta)\rangle$, and the overlap $\langle\Phi(\theta)|\Phi(\theta')\rangle=\mbox{Det}\left[\langle\psi_{j}(\theta)|\psi_{j'}(\theta')\rangle\right]$.
Then, $\nu$ obeys $e^{i\pi\nu}=\Pi_{\theta=0}^{\theta=2\pi-d\theta}\ \langle\Phi(\theta)|\Phi(\theta+d\theta)\rangle$
where $\Phi(2\pi)=\Phi(0)$. In practice, the matrix product $\Pi_{\theta=0}^{\theta=2\pi-d\theta}\langle\Phi(\theta)|\Phi(\theta+d\theta)\rangle$
is computed and the sum of the phases of its eigenvalues yields the
value of $\pi\nu$. According to the bulk boundary correspondence
principle, the winding number yields the number of SSs. In a clean
WNL, $\nu$ equals the number of drumhead states, as seen in Fig.~\ref{fig:wind-SS-all}(a)
for $W=0$ (where $L=12$ implies exactly $21$ drumhead states).

\begin{figure}[ht]
\centering{}\includegraphics[width=1.\columnwidth]{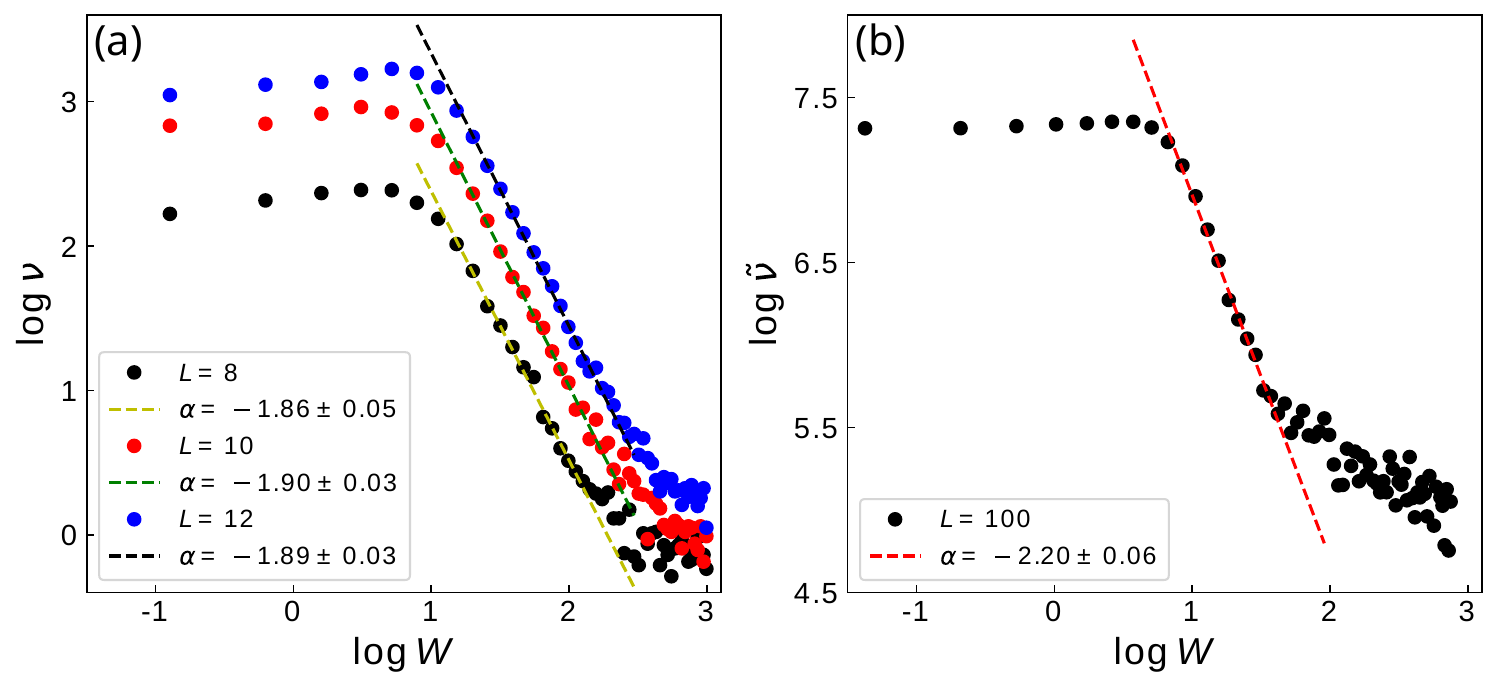}\caption{\label{fig:log-wind_log-SS}Log-log plots for the data shown in Fig.~\ref{fig:wind-SS-all}:
(a) disorder averaged winding number per unit area, $\nu/L^{2}$,
versus disorder strength, $W$, for clusters of $L^{3}$ cells; (b)
estimated number of low energy surface states, $\tilde{\nu}$, versus
$W$, obtained using $N_{m}=2^{10}$ polynomials for a system with
$100^{3}$ cells. Dashed lines are linear fits to the data.}
\end{figure}

The winding number as a function of increasing disorder is shown in
Fig.~\ref{fig:wind-SS-all}(a), where it can be seen to increase
with disorder for $W\lesssim2$. A transition to a topologically trivial
phase where $\nu=0$ is not observed. Instead, we find that the disorder
averaged winding number exhibits a power law decay $\nu\propto W^{\alpha}$,
with $\alpha\approx-1.9$ fairly insensitive to the system's size,
as illustrated in Fig.~\ref{fig:log-wind_log-SS}(a). The finite
winding number at finite chiral disorder is concomitant with a finite
bulk DOS at zero energy, as shown in the inset of Fig.~\ref{fig:wind-SS-all}(a)
alongside with the DOS for Anderson diagonal disorder, where a semimetal
to metal transition takes place at $W_{c}$ \citep{Goncalves2020}.
For any finite chiral disorder, the system is both topologically nontrivial
and a metal.

Counting the number of surface states as chiral disorder increases
would require treating exactly a semi-infinite system, thus avoiding
surface states hybridization on opposite boundaries. For the numerically
exact approach we have been following, surfaces are created in pairs,
so the presence of at least two is unavoidable. To minimize the mentioned
hybridization, we use a method which allows to reach large system
sizes: we study the change in the DOS that occurs when periodic boundary
conditions (PBC) along the $z$ direction are replaced by open boundary
conditions (OBC)~\citep{Silva2023}. Then, any change in the DOS
must be a surface effect. We define the density of surface states,
$\Delta\rho(E)\equiv\rho_{{\rm OBC}}(E)-\rho_{{\rm PBC}}(E)$, where
$\rho_{{\rm PBC}}$ ($\rho_{{\rm OBC}}$) denotes the DOS calculated
for PBC (OBC) along the $z$ direction. The integral of $\Delta\rho(E)$
over the whole energy axis vanishes because the total number of states
($2L^{3}$) is the same for OBC and PBC: bulk states are destroyed
to compensate for the creation of edge states. The number of low energy
surface states for OBC can be estimated by defining the energy interval,
$\left|E\right|<E_{w}$, around zero energy where $\Delta\rho(E)>0$.
Then, the integral 
\begin{equation}
\tilde{\nu}=L^{3}\int_{-E_{w}}^{E_{w}}\Delta\rho(E)dE\,,\label{DOSSint}
\end{equation}
provides an estimation of the number of low energy surface states.
We compute the DOS using the Kernel Polynomial Method (KPM) with an expansion
in Chebyshev polynomials to order $N_{m}$ \citep{Weisse2006}.

The results for $\tilde{\nu}$ are presented in Fig.~\ref{fig:wind-SS-all}(b)
and reveal an enhancement for $W\lesssim2$, in agreement with the
winding number in Fig.~\ref{fig:wind-SS-all}(a). For higher disorder,
$\tilde{\nu}$ decays as a power law, like the winding number $\nu$,
but with a different exponent $\alpha=-2.20\pm0.06$, which indicates
that $\tilde{\nu}<\nu$. In fact, we should expect $\tilde{\nu}$
to be a lower bound for the true number of drumhead states, even in
the absence of disorder: this is because the edge states near the
nodal line (where their localization length diverges) hybridize with
others on the opposite surface. The resulting energy splitting shifts
their energies to values higher than $E_{w}$, outside the integration
domain in Eq.$\,$(\ref{DOSSint}).

\begin{figure}[ht]
\centering{}\includegraphics[width=1.\columnwidth]{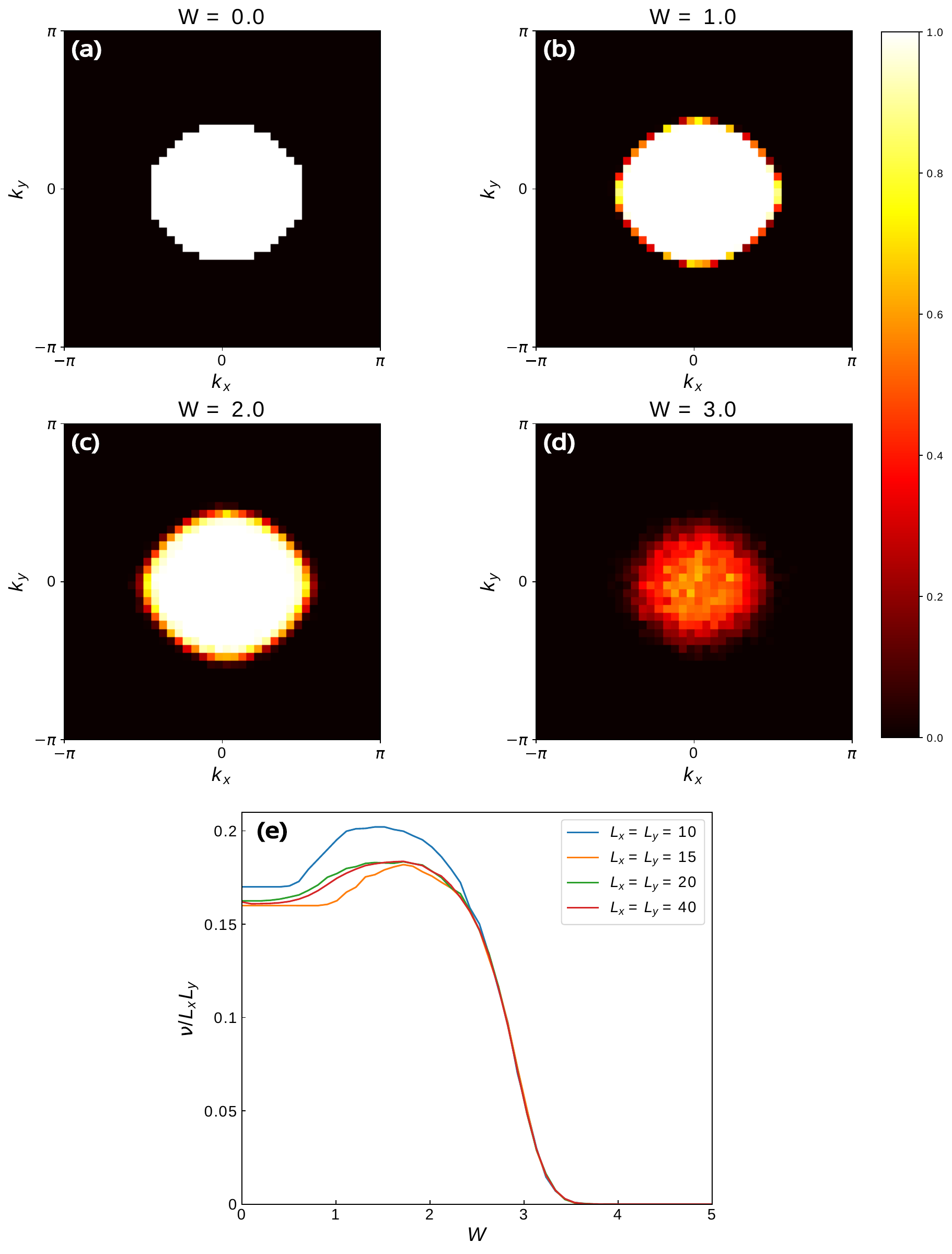}\caption{\label{fig:wind_prodan_plane}(a)-(d)~Momentum resolved winding number,
$\nu(k_{x},k_{y})$, for increasing disorder values $W$. (e)~Disorder
averaged total winding number per unit area versus disorder strength.
The system's length along $z$ is 100 unit cells.}
\end{figure}

The low disorder enhancement of surface states can be understood if
we consider the conceptually simpler case of disorder in one spatial direction,
where the random hopping terms only spatially depend on the $z$ coordinate,
by setting $\omega_{\bm{r}}^{x}=\omega_{\bm{r}}^{y}=\omega_{\bm{r}}^{0}=0$
in Eq.~(\ref{eq:disorder_V}). The system then behaves as a set of
decoupled chains along the $z$ axis, each labeled by $(k_{x},k_{y})$.
Each chain has a winding number, $\nu(k_{x},k_{y})$, which is unity
inside the nodal loop, and zero outside, in the clean case. Each chain
then undergoes a chiral disorder induced topological transition in
1D~\citep{Prodan2014}. Figure$\,$\ref{fig:wind_prodan_plane}(a)-(d)
shows the disorder averaged $\nu(k_{x},k_{y})$ at different disorder
values. It is seen that weak disorder slightly enlarges the area of
the central region. In this case, the total winding number $\nu$
is given by $\nu=\sum_{k_{x},k_{y}}\nu(k_{x},k_{y})$, and it is clearly
enhanced for $W\lesssim2$, as shown in Fig.$\,$\ref{fig:wind_prodan_plane}(e).
The physical interpretation is that the $(k_{x},k_{y})$ trivial chains
just outside the nodal line, which are close to a 1D topological transition,
become topological under weak disorder, as observed in topological
Anderson insulators \citep{Shen2009,Groth2009}. At stronger disorder,
$W\gtrsim2$, the chains gradually transit to the trivial phase, starting
from the periphery towards the inside. Figure$\,$\ref{fig:wind_prodan_plane}(e)
also shows that $\nu$ vanishes above a critical disorder strength,
as expected, signaling that all the chains (the 3D system) are topologically
trivial.

For the full 3D disordered case shown in Fig.~\ref{fig:wind-SS-all}(a),
no quantum states can be labeled by momentum. However, the enhancement
of $\nu$ at $W\lesssim2$ shows that some bulk quantum states are
close to becoming surface states upon adding weak disorder, a feature
shared with the case of disorder in one spatial direction. The difference to disorder in one spatial direction, here, is that no topological to trivial transition
is observed at high $W$. The degree of enhancement of $\nu / L^2$ can be increased or decreased by changing the model parameters.

\begin{figure}[ht]
\begin{centering}
\includegraphics[width=1.\columnwidth]{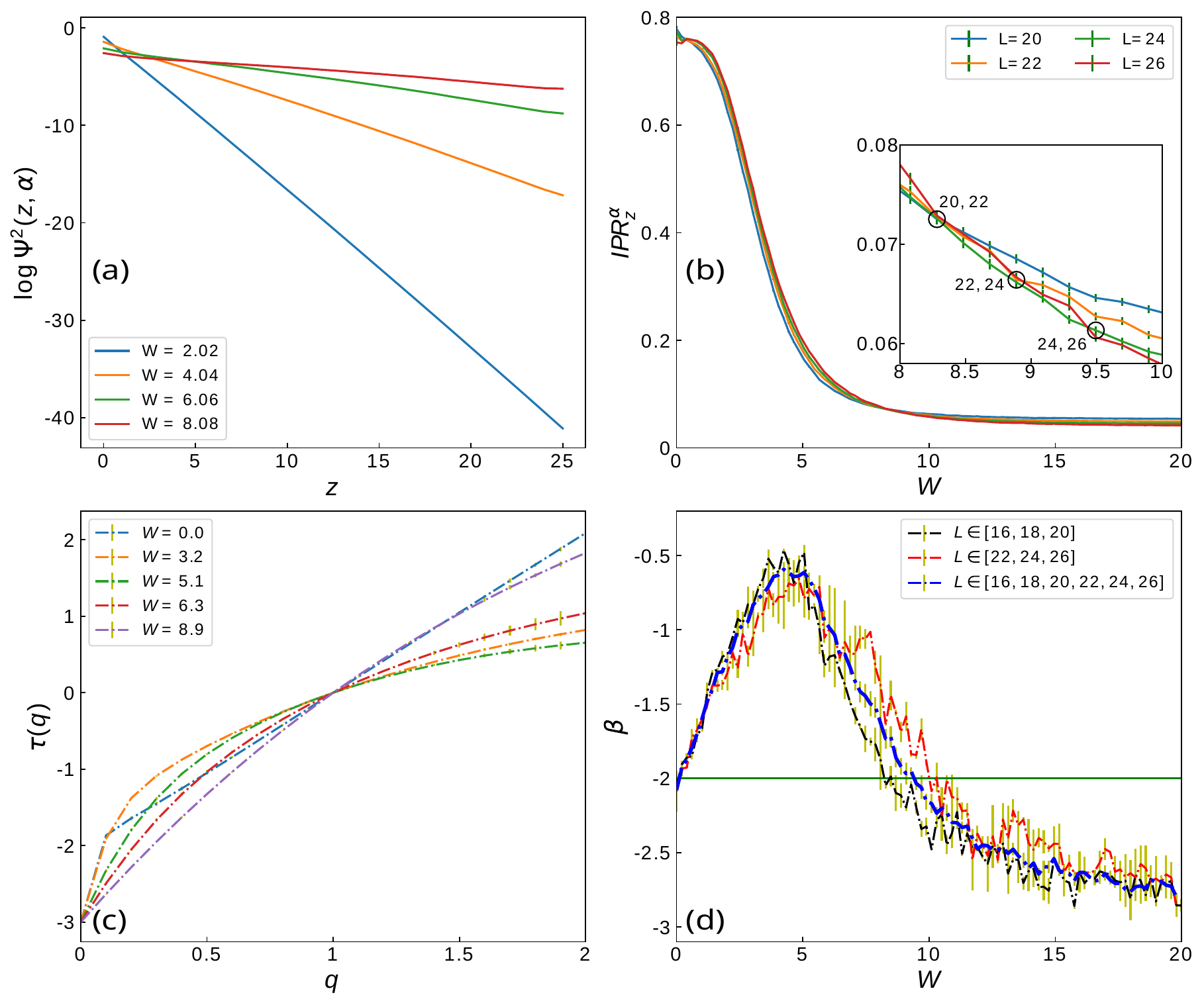}
\par\end{centering}
\centering{}\caption{\label{fig:ipr_z_alpha_prodan}Localization properties of the lowest
energy surface state: (a) exponential decay of probability, $|\Psi(z,\alpha)|^{2}$,
into the bulk for sublattice $\alpha=1$; (b) IPR$_{z}^{\alpha}$
versus disorder strength, $W$, for various system sizes, $L$; (c)
$\tau\left(q\right)$ for various $W$; (d) IPR scaling exponent,
$\beta=-\tau(2)$, versus $W$.}
\end{figure}

The presence of chiral symmetry and the existence of a winding number
imply that the associated surface states decay exponentially into
the bulk. In the clean limit, exponential localization of drumhead
states is well understood using, for example, the decomposition in
decoupled chains along $z$ mentioned above. In the following we demonstrate
that, for the chiral disordered system, localization of surface states
also takes place. The decay of probability into the bulk can be found
from an Inverse Participation Ratio (IPR) defined for the $z$ direction
in sublattice $\alpha$ as: 
\begin{equation}
{\rm IPR}_{z}^{\alpha}=\frac{\sum_{z}\Psi^{4}(z,\alpha)}{\left[\sum_{z}\Psi^{2}(z,\alpha)\right]^{2}}\,,\label{ipr_zalpha}
\end{equation}
where the probability $\Psi^{2}(z,\alpha)=\sum_{x,y}|\psi(x,y,z;\alpha)|^{2}$
is obtained from the surface state wave function $\psi(\bm{r},\alpha)$.
Figure~\ref{fig:ipr_z_alpha_prodan}(a) shows the exponential decay
of the lowest energy surface state probability from the $z=1$ surface.
Disorder \textit{increases} the localization length, $\xi$, in the
range of $W$ considered. This is confirmed in Fig.~\ref{fig:ipr_z_alpha_prodan}(b),
where IPR$_{z}^{\alpha}$ is a decreasing function of $W$. This behavior
is akin to the problem in 1D~\citep{Prodan2014}. It also explains
why $\tilde{\nu}$ decreases faster than $\nu$: since $\xi$ increases
with $W$, a larger fraction of surface states on opposite sides of
the sample significantly hybridize at larger $W$, falling out of
the integration energy window $E_{w}$ in Eq.$\,$\eqref{DOSSint}.
The inset to Fig.~\ref{fig:ipr_z_alpha_prodan}(b) shows that the
IPR$_{z}^{\alpha}$ monotonous dependence on $L$ inverts at $W\gtrsim8.5$.
At this disorder value, the localization length exceeds the system
sizes considered, pointing to a finite size effect. The inset also
shows that the crossing point between consecutive sizes, signaled
by the empty circles, shifts to higher disorder as $L$ increases,
in agreement with this scenario.

To gain a deeper understanding on the localization of surface states
we investigate their multifractal structure. This is done in real
space through the generalized IPR for a system with linear size $L$,
\begin{equation}
\mathcal{I}\left(q\right)=\frac{\sum_{\mathbf{r},\alpha}\left|\psi(\mathbf{r},\alpha)\right|^{2q}}{\left(\sum_{\mathbf{r},\alpha}\left|\psi(\mathbf{r},\alpha)\right|^{2}\right)^{q}}\propto L^{-\tau\left(q\right)}\,,\label{ipr_q}
\end{equation}
where $\psi(\mathbf{r},\alpha)$ is the lowest-energy surface state
amplitude at cell $\mathbf{r}$ and sublattice $\alpha$. Writing
the exponent as $\tau\left(q\right)=D\left(q\right)\cdot\left(q-1\right)$,
a constant $D(q)=D$ defines the fractal dimension of the wave function.
If $D(q)$ is not constant, the wave function is said to be multifractal
\citep{Janssen}. In Fig.~\ref{fig:ipr_z_alpha_prodan}(c) it can
be seen that $\tau(q)$ is non-linear for $W\neq0$, implying the
multifractality of surface states. Focusing on the $q=2$ case, the
quantity $\mathcal{I}\left(2\right)$ in Eq.~\eqref{ipr_q} becomes
the IPR. The scaling exponent IPR $\propto L^{\beta}$, with $\beta=-\tau(2)$,
is shown in Fig.~\ref{fig:ipr_z_alpha_prodan}(d) as a function of
the disorder strength $W$. For the $W=0$ (clean limit), we obtain
the exponent $\beta=-2$ as expected for a surface state that extends
in the $xy$ surface and is localized along $z$. However, for $W\neq0$,
the exponent reveals a highly nontrivial fractal dimension dependence
on disorder.

For high disorder, $W\gtrsim10$, the slope of $\tau(q)$ becomes
approximately 3 ($D(q)\approx3$), indicating that the probability
occupies the full three-dimensional volume. This is already apparent
for $W=8.9$ in Fig.~\ref{fig:ipr_z_alpha_prodan}(c), particularly
at low $q$. We believe that this is because the localization length
exceeds $L$ and that this behavior breaks down for large enough systems.
The value $|\beta|>2$ for $W\gtrsim8.5$ in Fig.~\ref{fig:ipr_z_alpha_prodan}(d)
is also likely due to $\xi$ exceeding the system sizes considered.
Notice that in Fig.~\ref{fig:ipr_z_alpha_prodan}(d) the $W$ value
at which $|\beta|\approx2$ increases if we consider only the largest
sizes. Other low energy states present the same features shown in Fig.~\ref{fig:ipr_z_alpha_prodan}.

\begin{figure}[ht]
\centering{}\includegraphics[width=1.\columnwidth]{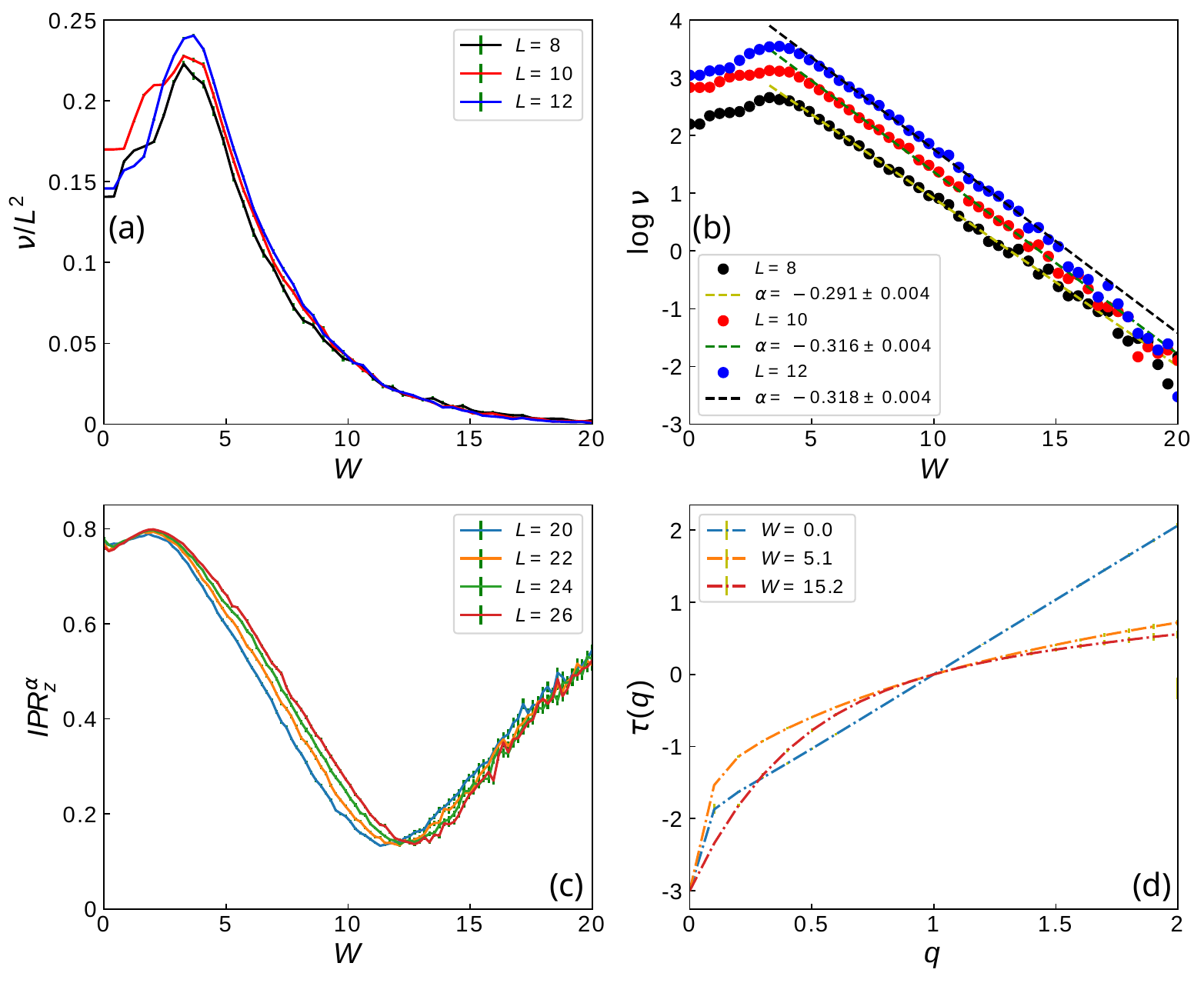} \caption{\label{fig:wind_m} (a) Disorder averaged winding number per unit
area, $\nu/L^{2}$, versus intra-cell disorder $W$ for clusters of
$L^{3}$ sites. (b) Log-linear plot for the data in (a). (c) IPR$_{z}^{\alpha}$
versus disorder strength, $W$, for various system sizes, $L$. (d) $\tau\left(q\right)$
for various $W$.}
\end{figure}

We now turn to the case where only the intra-cell hopping term, $m$,
is disordered ($\omega_{\bm{r}}^{\delta}=0$, except if $\delta=0$).
Two qualitative differences to the previous fully disordered case are
found: an exponential decrease of $\nu$ with disorder strength, as
seen in Fig~\ref{fig:wind_m}(a) and in~\ref{fig:wind_m}(b), with
similar results for $\tilde{\nu}$ (not shown); and Fig~\ref{fig:wind_m}(c) suggests an Anderson localization
transition at higher $W$~\citep{Luo2020}, which we have confirmed with the Transfer Matrix Method (not shown) \citep{PhysRevLett.131.056301}, and does not exist
in the previous case. We find that for a disorder strength $15 \lesssim W \lesssim 18$ the system is in a quasilocalized phase \citep{PhysRevLett.131.056301,zhao2024topologicaleffectandersontransition}. However, unlike \citep{PhysRevLett.131.056301}, we do not see the winding number vanish after the Anderson transition, which, we believe, is due to finite size effects. Since, in Fig~\ref{fig:ipr_z_alpha_prodan}(b), we do not find an Anderson transition, we believe that a quasi-localized phase is not present in the fully disordered model.

The dependence of the localization length of surface states on disorder
is also more complex. Fig~\ref{fig:wind_m}(c) for IPR$_{z}^{\alpha}$
shows that $\xi$ first decreases ($W\lesssim2$), then increases
($2\lesssim W\lesssim12$) and finally decreases beyond $W\gtrsim12$, indicative of proximity to an Anderson transition. The exponent $\tau(q)$, shown in Fig~\ref{fig:wind_m}(d),
exhibits multifractal behavior.

In summary, we have established the fate of a WNL semimetal under
the presence of chiral disorder and shown that the semimetallic phase
is unstable to a topological metal. The coexistence of topological
surface and bulk extended states of the disordered metal is a consequence
of the finite winding number in agreement with the bulk-edge correspondence
principle. The surface states are robust up to very large disorder,
decaying exponentially into the bulk in the direction orthogonal to
the nodal loop. They exhibit multifractal properties along the surface
that strongly and nontrivially depend on disorder strength, thus realizing
a different type of bound states in the continuum \citep{Yang2013,Hsu2016,bic_PhysRevLett.118.166803,bic_PhysRevB.100.075120,Benalcazar2020}.
To our knowledge, this is the first example of a 3D intrinsically
disordered topological metal. The observable signatures of this exotic
state of matter, such as transport properties, should be investigated
in the future. 
\begin{acknowledgments}
The authors MG, PR and MA acknowledge partial support from Fundação
para a Ciência e Tecnologia (FCT-Portugal) through Grant No. UID/CTM/04540/2019.
MG and EVC acknowledge partial support from FCT-Portugal through Grant
No. UIDB/04650/2020. MG acknowledges further support from FCT-Portugal
through the Grant SFRH/BD/145152/2019. We finally acknowledge the
Tianhe-2JK cluster at the Beijing Computational Science Research Center
(CSRC) and the OBLIVION supercomputer (based at the High Performance
Computing Center - University of Évora) funded by the ENGAGE SKA Research
Infrastructure (reference POCI-01-0145-FEDER-022217 - COMPETE 2020
and the Foundation for Science and Technology, Portugal) and by the
BigData@UE project (reference ALT20-03-0246-FEDER-000033 - FEDER)
and the Alentejo 2020 Regional Operational Program. Computer assistance
was provided by CSRC, CENTRA/IST and the OBLIVION support team. 
\end{acknowledgments}

\bibliography{NodalLoops}

\end{document}